%% file: main.tex
\documentclass[a4paper]{article}
\usepackage{INTERSPEECH2021}
\input{inputfiles/packages}
\pgfplotsset{compat=1.17}

\title{End-to-End Open Vocabulary Keyword Search}
\name{Bolaji Yusuf$^{1,2}$, Alican Gok$^1$, Batuhan Gundogdu$^{1,3}$, Murat Saraclar$^1$\thanks{The work was partly supported by European Union’s Horizon 2020 project No. 870930 - WELCOME, Boğaziçi University Research Fund under the Grant Number 16903, and the Turkish Directorate of Strategy and Budget under the TAM Project number 2007K12-873.}}
\address{
  $^{1}$Bogazici University, Department of Electrical and Electronics Engineering, Turkey\\
  $^2$Brno University of Technology, Faculty of Information Technology, Speech@FIT, Czechia\\
  $^3$University of Chicago, Department of Radiology, IL, USA}
\email{$\{$bolaji.yusuf,alican.gok,batuhan.gundogdu,murat.saraclar$\}$@boun.edu.tr}
\begin{document}

\maketitle
\begin{abstract}
 \input{inputfiles/abstract}
\end{abstract}


\section{Introduction}
\label{sec:introduction}
\input{inputfiles/intro}

\input{inputfiles/related}
\vspace{-0.05in}
\section{Model}
\label{sec:model}
\input{inputfiles/model}
\vspace{-0.03in}
\section{Experiments}
\label{sec:experiments}
\input{inputfiles/experiments}

\vspace{-0.05in}
\section{Conclusion}
\input{inputfiles/proclusion}

\bibliographystyle{IEEEtran}

\bibliography{mybib}

\end{document}

%% file: inputfiles/packages.tex
\usepackage{amsmath,graphicx}
\usepackage{amssymb}
\usepackage{bbm}
\usepackage{textcomp}
\usepackage{mathtools}
\usepackage{booktabs}
\usepackage[utf8]{inputenc}
\usepackage[dvipsnames]{xcolor}
\usepackage{mystyle}
\usepackage{multirow}
\usepackage{tikz}
\usepackage{hyperref}
\usepackage{pgfplots}
\usepackage{tikz-3dplot}
\usetikzlibrary{arrows.meta}
\usetikzlibrary{positioning,through, backgrounds,decorations.pathreplacing,shapes.misc}

\tdplotsetmaincoords{70}{30}

%% file: inputfiles/abstract.tex
Recently, neural approaches to spoken content retrieval have become popular. However, they tend to be restricted in their vocabulary or in their ability to deal with imbalanced test settings. These restrictions limit their applicability in keyword search, where the set of queries is not known beforehand, and where the system should return not just whether an utterance contains a query but the exact location of any such occurrences. In this work, we propose a model directly optimized for keyword search. The model takes a query and an utterance as input and returns a sequence of probabilities for each frame of the utterance of the query having occurred in that frame. Experiments show that the proposed model not only outperforms similar end-to-end models on a task where the ratio of positive and negative trials is artificially balanced, but it is also able to deal with the far more challenging task of keyword search with its inherent imbalance. Furthermore, using our system to rescore the outputs an LVCSR-based keyword search system leads to significant improvements on the latter.

%% file: inputfiles/intro.tex
Many technologies have emerged to facilitate browsing the vast amounts of spoken content that have become commonplace. One of these is keyword search (KWS), in which a user-specified keyword (also referred to as query or term) is searched within an audio archive (or document), and the locations, if any, are returned along with confidence scores.

The traditional approach to KWS entails using an LVCSR system to transcribe the archive, and then conducting text retrieval on the LVCSR output~\cite{can2011lattice}. This approach has been successfully applied for retrieval on broadcast news~\cite{wang2008comparison,parlak2012performance}, video lectures~\cite{garcia2006keyword} and web videos~\cite{lee2005combining}.

A limitation of the LVCSR approach is the inability to retrieve words that are not seen at training time. Retrieving such out-of-vocabulary (OOV) terms requires extra engineering. The most common approach to open-vocabulary search is to use subword units instead of words as the ASR output, thereby exchanging lattice compactness for coverage~\cite{mamou2007vocabulary,karakos2014subword,szoke2008hybrid}. Another approach is to use phone-confusion statistics to convert an OOV term into a similar sounding in-vocabulary (IV) one and then searching for this \textit{proxy} term instead~\cite{saraclar2013confusion,chen2013using}. More recently, dynamic time warping (DTW)-based methods which completely eschew LVCSR have been shown to outperform both~\cite{gundougdu2017joint,yusuf2019low}.

%% file: inputfiles/related.tex
Recently, several neural architectures have been proposed for hot-word spotting. In~\cite{chen2014small}, a neural network is used to predict keywords from spectral features. CNN-based systems have also been applied for supervised~\cite{Segal2019} and weakly-supervised ~\cite{palaz2016jointly} keyword spotting. 
In~\cite{alvarez2019end}, a system based on rank-1 encoder-decoder networks was proposed for end-to-end streaming keyword spotting. These systems have the limitation of requiring a pre-specified set of keywords.

End-to-end networks that bypass LVCSR while also being able to deal with open-vocabulary applications have also been studied. In~\cite{audhkhasi2017end}, a pair of auto-encoders are pre-trained to encode the query and document, and a feed-forward network is used to predict whether the query encoding occurs in the document encoding. Other similar works use acoustic word embeddings for metric-based keyword search~\cite{chen2019audio,kamper2020improved}. In~\cite{zhao2020end}, the authors improve on~\cite{audhkhasi2017end} by improving the pre-training and also by leveraging temporal information available in the form of forced-alignments. While these methods have been shown to perform well for well-balanced tasks such as classifying whether two embeddings belong to the same term or whether an utterance contains a query, they have limited applicability for actual keyword search, which is extremely imbalanced in that most locations do \textit{not} contain the keyword; for every positive trial, there are thousands of negative ones.

In this paper, we propose an end-to-end keyword search model that takes a query as input and returns frame-level probabilities of the query occurring in the document. Unlike other neural KWS architectures, we do not need to construct positive-negative samples for testing but can directly search on whole utterances. We show that our model outperforms other end-to-end KWS systems in segment-level classification, performs well in realistic KWS settings, and is able to improve the performance of a competitive LVCSR-based system by rescoring.

%% file: inputfiles/model.tex
We formulate keyword search as the task of classifying whether a keyword occurs at any given location in the document. This formulation -- as opposed to, say, a regression task predicting the boundaries of a hypothesized hit -- is very well suited to being used in combination with other keyword search systems since it allows us to get a keyword search result for any possible subsequence of the document.

Given a query phrase $\Matrix{q}$ and an utterance represented as a sequence of frames $\Matrix{X} = \bigl(\Matrix{x}_1, \dots, \Matrix{x}_N\bigr)$, we seek the sequence $\Matrix{y}{(\Matrix{q}, \Matrix{X})}= (y_1, \dots, y_{N}) \in \{0, 1\}^N$ such that:
\vspace{-0.05in}
\begin{align}
    y_n=
    \begin{cases}
      1, & \text{if}\ \Matrix{q} \text{ occurs at frame } n \text{ of } $\Matrix{X}$ \\
      0, & \text{otherwise}.
    \end{cases}
\end{align}
Note that although we elect not to write it explicitly to reduce clutter, each $y_n$ is a function of $\Matrix{q}$ and $\Matrix{X}$. We can then train a neural network with parameters $\Matrix{\theta}$ to obtain:
\begin{align}
    \Matrix{\theta}^* = \argmax_{\Matrix{\theta}} \sum_{ \Matrix{q}} \sum_{\Matrix{X}} \sum_n \log p_{\Matrix{\theta}} ( y_n | \Matrix{q}, \Matrix{X}),
    \label{eq:posterior}
\end{align}
where the summations are over all the phrases, utterances and time frames in the training set.

\subsection{Model definition}
\begin{figure}[t]
    \centering
    \hspace*{-0.37in}
    \includegraphics[width=1.12\linewidth]{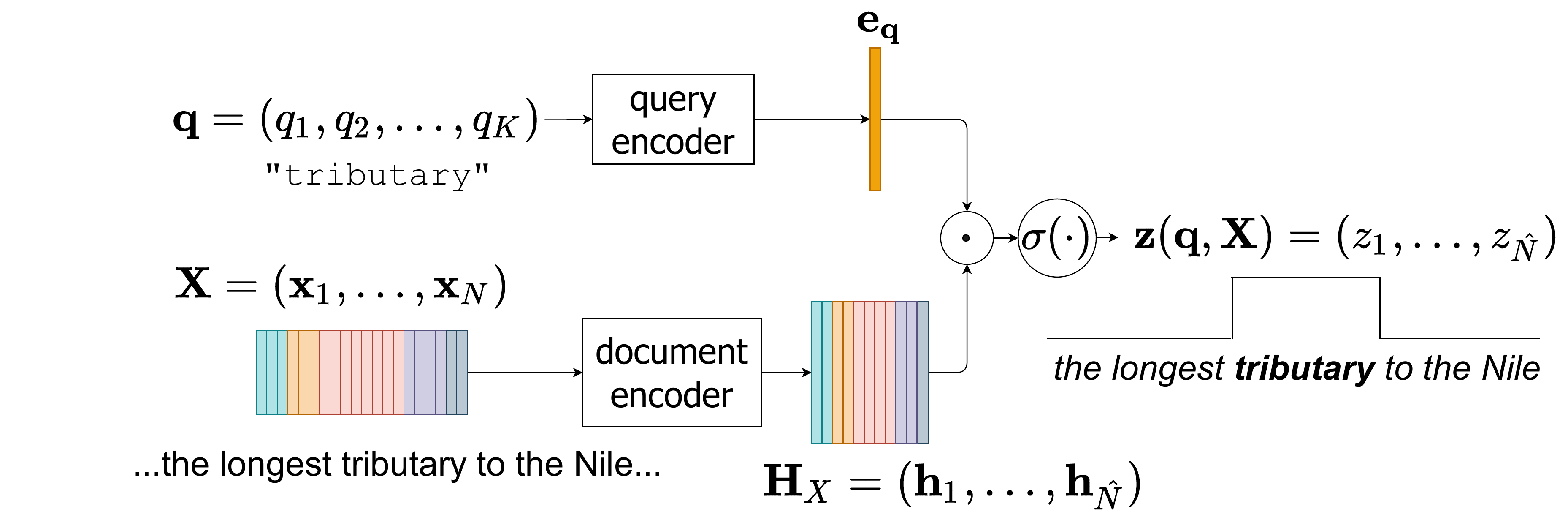}
    \caption{Overview of the proposed model.}
    \label{fig:model}
    \vspace{-0.15in}
\end{figure}

Our keyword search model, depicted in Figure~\ref{fig:model}, comprises a recurrent document encoder and a recurrent query encoder. We conduct the search via a matrix-vector multiplication of the encoder outputs which results in a vector of logits for each time frame. The logistic sigmoid is then used to compute the posterior probabilities $p_{\Matrix{\theta}}(y_n|\dots)$ which we post-process to detect the locations of each keyword.

\subsubsection{Query encoder}
The input to the query encoder is a sequence of grapheme indices $\Matrix{q} = (q_1, \dots, q_K)$ that constitute the query and the output is a fixed-length representation $\Matrix{e}_{\Matrix{q}} \in \mathbb{R}^D$. A trainable embedding layer converts the sequence of letters into a sequence of vectors which are input into a stack of bidirectional gated recurrent unit (GRU) layers. The GRU outputs another sequence of vectors $\Matrix{V} = (\Matrix{v}_1, \dots, \Matrix{v}_K)$. The final query representation is then computed from the sum of these GRU output vectors along the sequence axis thus:
\vspace{-0.04in}
\begin{align}
    \Matrix{e}_{\Matrix{q}} = \Matrix{W}_1 (\sum_{k=1}^{K} \Matrix{v}_k) + \Matrix{b_1},
    \label{eq:query_rep}
\end{align}
where $\Matrix{W}_1$ and $\Matrix{b}_1$ are the weight and bias of an affine transform that changes the dimensionality of the query representation to ensure it matches the output of the document encoder.

We experimented with using the output of the GRU at final frame ($\Matrix{v}_K$) instead of the summation in~\eqref{eq:query_rep}, but found that the summation performed better. We also experimented with having a unidirectional query encoder but we found the bidirectional encoder to be better.

\subsubsection{Document encoder}
The input to the document encoder is the sequence of speech features $\Matrix{X}$ of length $N$. First, $\Matrix{X}$ is passed through a stack of BiLSTM layers which output $\Matrix{U} = (\Matrix{u}_1, \dots, \Matrix{u}_{\hat{N}})$ of length $\hat{N}$. The final encoder output $\Matrix{H}_{\Matrix{X}} \in \mathbb{R}^{\hat{N} \times D}$ is then:
\vspace{-0.04in}
\begin{align}
    \Matrix{H}_{\Matrix{X}} = \Matrix{W}_2 \Matrix{U} + \Matrix{b}_2,
\end{align}
where $\Matrix{W}_2$ and $\Matrix{b}_2$ constitute an affine transformation similar to that at the output of the query encoder, and the addition is done by broadcasting $\Matrix{b}_2$ across the temporal axis.

We down-sample the hidden representations between some of the BiLSTM layers (so that $\hat{N} = \floor{\frac{N}{s}}$ for some $s$). This decreases the computational cost of the search, and we found empirically that it improves the search accuracy.

\subsubsection{Search function}
The search output is computed by taking a matrix-vector product of the encoder outputs, followed by a logistic sigmoid to compute the desired vector of per-frame posterior probabilities $\Matrix{z}(\Matrix{q}, \Matrix{X})=(z_1, \dots, z_{\hat{N}}) \in (0, 1)^{\hat{N}}$ thus:
\vspace{-0.04in}
\begin{align}
    \Matrix{z}(\Matrix{q}, \Matrix{X}) = \sigma(\Matrix{H}_{\Matrix{X}} \Matrix{e}_{\Matrix{q}}).
    \label{eq:output}
\end{align}
Note that the document and query representations only interact through this product and are otherwise independent. Therefore, we can amortize the cost of search by pre-computing the document representation. For each query, we can then compute its representation and take the product.

\subsection{Model training}
Our training procedure involves optimizing a modified form of~\eqref{eq:posterior}. The first two summations~\eqref{eq:posterior} are over all phrases and utterances in the training set. For a corpus of $\mathcal{U}$ utterances with $\mathcal{W}$ words each, there are $\mathcal{O}(\mathcal{U} \mathcal{W}^2)$ elements in the double summation. Therefore, we make the following modifications to reduce the training cost:
\begin{enumerate}
    \item We limit training phrases to unigrams, bigrams and trigrams.
    \item At each training step, we sample a batch $\mathcal{Q}$ of such phrases for the first summation instead of the whole set.
    \item For each training phrase $\Matrix{q} \in \mathcal{Q}$, we sample $\mathcal{X}_{\Matrix{q}}$, a set of $\mathcal{M} \ll \mathcal{U}$ utterances, at least one of which contains $\Matrix{q}$.
\end{enumerate}
When sampling, we consider different examples of the same phrase to be different tokens, so that more frequently occurring phrases are naturally sampled more often. The utterances $\mathcal{X}_{\Matrix{q}}$ are re-sampled at each training step.

For each query-utterance training pair $(\Matrix{q}, \Matrix{X})$, we define a loss function between the sigmoid outputs $\Matrix{z}(\Matrix{q}, \Matrix{X})$ and the (down-sampled) labels $\Matrix{y}(\Matrix{q}, \Matrix{X})$:
\vspace{-0.04in}
\begin{align}
    J_s(\Matrix{q}, \Matrix{X}) = -\sum_{n=1}^{\hat{N}}& \Bigl (\mathbbm{1}_{z_n > 1 - \phi} \cdot (1-y_{n}) \log (1-z_n)
     \nonumber \\
    & + \mathbbm{1}_{z_n < \phi} \cdot \lambda \cdot y_{n} \log z_n \Bigr).
    \label{eq:loss_single}
\end{align}
$\lambda$ is a hyper-parameter that controls the relative weight of frames labeled $1$ to frames labeled $0$. $\phi$ controls the sensitivity of the loss function to easily classified frames: frames labeled $1$ with sigmoid outputs already above $\phi$ and frames labeled $0$ with sigmoid outputs already below $1-\phi$ do not contribute to the loss function as they are considered good-enough. This prevents the model from learning to better classify frames that are already well classified at the expense of learning to classify difficult frames. Observe that if we set $\lambda=1$ and $\phi = 1$, then the loss function returns to the binary cross-entropy function.

\subsection{Post-processing for search}
Having obtained the vector of probabilities from~\eqref{eq:output}, we post-process them to perform various keyword search tasks.
\subsubsection{Segment classification}
\label{subsec:disc}
This task is a simplified form of keyword search where it is enough to classify whether or not a speech segment contains the given query. To do this, we simply compute a segment-wide score:
    $p = \max_n z_n,
    \label{eq:sur}$
and say that the segment contains the query if $p$ is larger than some threshold, which is tuned on a development set. The rationale for this method is that it is enough for a query to occur at any frame of a segment for it to have occurred in that segment. Therefore, we need only care about the frame with the highest probability.

\subsubsection{Keyword search}
\label{subsec:standalone_kws}
\begin{figure}[t!]
    \centering
    \includegraphics[width=5cm]{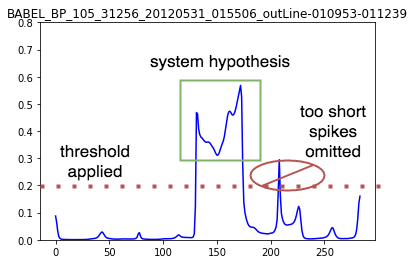}
    \caption{System hypothesis calculation from network outputs.}
    \label{fig:golden_detection}
    \vspace{-0.15in}
\end{figure}

The task here is to return a location within the utterance hypothesized to contain the query along with a confidence score. The procedure, illustrated in Figure~\ref{fig:golden_detection}, is as follows:
\begin{enumerate}
    \item We set the probabilities ($z_n$) below some threshold to zero. This threshold is tuned on the development set.
    \item We pick the ``islands" of non-zero elements as our system hypotheses. The confidence score is computed as the median probability of each interval. We also experimented with the mean and max operations but found median to be better.
\end{enumerate}
The thresholding is a necessary first step because the sigmoid outputs strictly positive values. Additionally, we found that pruning out intervals which are shorter than $20$ms $\times$ \#letters for each query led to slightly improved performance.

\vspace{-0.03in}
\subsubsection{Keyword search rescoring}
\label{subsec:rescoring}
In this task, we seek to improve the performance of another keyword search system by fusion of scores. Our approach entails rescoring the hypothesis list of a baseline system. Our proposed model is particularly well suited to this task since it outputs scores for each frame, and so we can readily obtain a score for any location with which we are presented. For a given query-document pair, if the baseline hypothesizes a start-end-score triplet $(n_1, n_2, \tilde{p})$, we return a new triplet $(n_1, n_2, p)$ at the same location with score given by a weighted sum of the original score and the average of the network's outputs in that interval, $p = \gamma \tilde{p} + \text{\texttt {mean}}( \Matrix{z}[n_1:n_2])$, 
where $\gamma$ is a weighting hyper-parameter which we tune on the development set. We experimented with using the median and max instead of the mean but found no significant improvements.

%% file: inputfiles/experiments.tex
\subsection{Experiment setup}
\subsubsection{Data}

We experiment on the limited language pack (LLP) data from the IARPA Babel program~\cite{harper2014iarpa}. For each language, this comprises a 10-hour training set used to train a speaker-adapted HMM-GMM model with which we obtain word level training alignments, a 10-hour development set to tune model hyper-parameters, a 5-hour ``evalpart1" set for evaluation, and keyword lists whose distributions are shown in Table~\ref{tab:kwdist}.

We consider two kinds of acoustic features as input to the document encoder: 80-dimensional filter-banks and 42-dimensional multilingual bottleneck features (BNF), each with frames of length 10ms. The BNF extractor is trained with 19 languages' LLPs (totaling about 190 hours) including Assamese and Bengali but not Pashto, Turkish and Zulu.
\vspace{-0.03in}
\subsubsection{Network configuration}
\begin{table}[t]
    \def\arraystretch{0.8}
    \setlength\tabcolsep{4pt}
    \centering
    \caption{Number of dev and eval queries in each language.}
    \label{tab:kwdist}
    \begin{tabular}{lcccccc}
    \toprule
    Language & Assamese & Bengali & Pashto & Turkish & Zulu \\
    \midrule
    Dev-IV  & 1403 & 1295 & 1465 & 219 & 1194 \\
    Dev-OOV  & 597 & 705 & 600 & 88 & 806 \\
    Eval-IV & 5286 & 4957 & 3233 & 1955 & 2199 \\
    Eval-OOV & 2087 & 2368 & 970 & 1216 & 1111 \\
    \bottomrule
    \vspace{-0.3in}
    \end{tabular}
\end{table}
The document encoder consists of 6 BiLSTM layers each with 512 dimensional output in each direction. Each BiLSTM layer (including the first) is preceded by a batch-normalization layer and followed by a dropout layer with dropout probability of 0.4. In addition, the outputs of the first and fourth BiLSTM layers are each down-sampled by a factor of 2 (i.e. $\hat{N} = \floor{\frac{N}{4}}$). The final affine projection has an output dimension of $D=400$.

The query encoder is composed of a 32-dimensional embedding layer, two 256-output-dimensional BiGRU layers each preceded by a batch-normalization layer and a final affine layer with 400-dimensional output.

We train with Adam using an initial learning rate of $2\times10^{-4}$. We use a batch size of $|\mathcal{Q}| = 64$ and $|\mathcal{X}|_{\Matrix{q}} = 4$ utterances per training phrase. The training hyper-parameters in~\eqref{eq:loss_single} are empirically chosen as $\lambda=5$ and $\phi = 0.7$. We select a random 10\% of the training utterances for validation. We halve the learning rate whenever the validation loss stagnates for 4 epochs and halt training when it stagnates for 10 epochs.

\subsection{Classification task}
Our first experiment involves classifying whether a given segment of speech contains a keyword. This task serves as a benchmark for comparing our approach with recent end-to-end keyword search sytems~\cite{audhkhasi2017end,zhao2020end}. We use the evaluation setup from~\cite{zhao2020end}. Each utterance is divided into one-second segments with 50\% overlap. Any segment that contains even a portion of the keyword is considered a positive test for that keyword. For example, a query which occurs between the 0.4-1.2 second marks of some utterance will yield three positive tests (the ranges 0-1, 0.5-1.5 and 1-2). For each such positive test, a random negative test (3 in total for the given example) is generated from other locations (possibly other utterances) which do not contain the query. Each test returns a score which should be high for positive and low for negative tests. The performance is measured in terms of accuracy and area under the curve (AUC).

We take the systems from~\cite{audhkhasi2017end} and~\cite{zhao2020end} as baselines. In both cases, we directly take the results reported in~\cite{zhao2020end}. For our system, we use the procedure described in Section~\ref{subsec:disc} with threshold determined on the dev-set.

Table~\ref{tab:acc_auc} shows the results obtained. Across all languages, our system gives significant improvements in both accuracy and AUC over the baselines. We get still further improvements by switching from filter-banks to BNF. We report P-B results for Assamese and Bengali for completeness, but keep in mind that the BNF extractor training includes those languages.

\begin{table}[t!]
    \def\arraystretch{0.9}
    \centering
    \caption{Accuracy and AUC on the eval set. B1 and B2 refer to the baseline systems from~\cite{audhkhasi2017end} and~\cite{zhao2020end} respectively. P-F and P-B refer to our system using filter-banks and bottleneck features respectively.}
    \vspace{-0.05in}
    \begin{tabular}{llccccc}
    \toprule
        && \multicolumn{2}{c}{Accuracy} && \multicolumn{2}{c}{AUC}
\\
         Language & System & IV & OOV && IV & OOV \\
         \midrule
         Assamese & B1 & 0.604 & 0.604 && 0.638 & 0.632 \\
         &B2& 0.626 & 0.619 && 0.724 & 0.724 \\
         &P-F & \textbf{0.713} & \textbf{0.677} && \textbf{0.797} & \textbf{0.778} \\
         \cmidrule{2-7}
         &P-B & \textbf{0.824} & \textbf{0.811} && \textbf{0.909} & \textbf{0.896} \\
         \midrule

         Bengali & B1 & 0.589 & 0.579 && 0.658 & 0.653 \\
         &B2 & 0.639 & 0.637 && 0.743 & 0.742 \\
         &P-F & \textbf{0.744} & \textbf{0.696} && \textbf{0.840} & \textbf{0.769} \\
         \cmidrule{2-7}
         &P-B & \textbf{0.836} & \textbf{0.784} && \textbf{0.912} & \textbf{0.871} \\
         \midrule

         Pashto & B1 & 0.608 & 0.594 && 0.653 & 0.634 \\
         &B2& 0.674 & 0.665 && 0.797 & 0.786 \\
         &P-F & \textbf{0.857} & \textbf{0.830} && \textbf{0.927} & \textbf{0.909} \\
         \cmidrule{2-7}
         &P-B & \textbf{0.881} & \textbf{0.855} && \textbf{0.946} & \textbf{0.925} \\
    \bottomrule
    \end{tabular}
    \label{tab:acc_auc}
    \vspace{-0.1in}
\end{table}

\begin{figure}[t]
    \centering
    \begin{tikzpicture}
        \input{figures/ablation}
    \end{tikzpicture}
    \caption{Eval-set ATWV of various ablations. Every other bar is obtained by changing one component from the \texttt{base} system.}
    \label{fig:standalone}
    \vspace{-0.2in}
\end{figure}
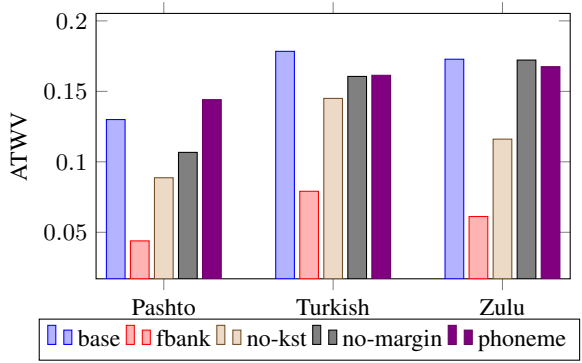

\subsection{Keyword search}
In this set of experiments, the task is to return the exact locations of each query within utterances along with the confidence scores. Unlike the balanced classification task, here the problem is made more complex by the presence of more negative trials than positive ones. For instance, the Pashto dev-set contains 14451 examples and 72 million negative trials keeping the rate of one trial/second, which makes metrics like accuracy unsuitable. Therefore, we use the term weighted value (TWV) metric proposed by NIST~\cite{Fiscus2006} for KWS evaluations~\cite{harper2014iarpa,Fiscus2007,kws14Evalplan,byers20192017}. The TWV for a set of queries $\mathcal{Q}$ and a threshold $\theta$ is computed as follows:
\vspace{-0.03in}
\begin{equation}
TWV(\theta, \mathcal{Q}) = 1-\frac{1}{|\mathcal{Q}|}\sum_{q \in \mathcal{Q}}(P_{miss}(q,\theta) + \beta P_{FA}(q,\theta)) ,
\label{eqn:twv}
\end{equation}

where $P_{miss}(q,\theta)$ and $P_{FA}(q,\theta))$ are the probabilities of misses and false alarms respectively, and $\beta$ controls the relative costs of false alarms and misses. As in the NIST STD evaluations~\cite{Fiscus2007}, we set $\beta=999.9$. 
On the dev-set, we report the maximum TWV (MTWV) which is the TWV at the threshold that maximizes it. This threshold is fixed and used to compute the actual TWV (ATWV) for the eval-set. Since TWV requires a single threshold, it is necessary to normalize the scores across keywords. We adopt keyword specific thresholding (KST) from~\cite{miller2007rapid} for score normalization.

\begin{table}[t]
    \def\arraystretch{0.8}
    \centering
    \caption{Results of rescoring LVCSR output. L denotes the LVCSR baseline. L $\oplus$ P denotes the result of rescoring.}
    \begin{tabular}{llcccc}
    \toprule
         Language & System & Dev MTWV & Eval ATWV\\
         \midrule
         Pashto & L & 0.2754 & 0.3190 \\
         &L $\oplus$ P& \textbf{0.3068} & \textbf{0.3464} \\ 
         \midrule
         
         Turkish & L & 0.4892 & 0.4051 \\
         &L $\oplus$ P&\textbf{0.5080}& \textbf{0.4419} \\
         \midrule
         
         Zulu & L & 0.3716 & 0.3381 \\
         &L $\oplus$ P& \textbf{0.3917}& \textbf{0.3620}\\
    \bottomrule
    \end{tabular}
    \label{tab:fusion}
    \vspace{-0.2in}
\end{table}

As stated before, Assamese and Bengali are included in the BNF extractor training, so we elect not to report results for those. Instead we use Turkish and Zulu, which, along with Pashto, are not included in the BNF training.

Figure~\ref{fig:standalone} shows the performance of our system when used as a standalone KWS system by following the procedure described in Section~\ref{subsec:standalone_kws} with KST normalization (\texttt{base}).
{For context, our re-implementation of~\cite{audhkhasi2017end} (using BNF) achieved MTWV of 0 and negative ATWV.} The figure also shows the impact of changing various components of our system.
First, we find that, as expected in this low-resource setting, filter-banks (\texttt{fbank}) perform worse than BNF. We also observe deterioration in performance when we omit score normalization (\texttt{no-kst}).

Next, we measured the impact of removing the margin term from our loss function by setting $\phi=1$ in ~\eqref{eq:loss_single} which results in a weighted binary cross-entropy objective (\texttt{no-margin}). For Turkish and Pashto, we find that the margin term helps significantly, whereas for Zulu, the impact is milder.

Furthermore, we considered the impact of using  phonemic instead of graphemic query representation (\texttt{phoneme}). When using phonemes, we use the lexicon available with the Babel data for training and for IV terms, and we train a G2P model~\cite{bisani2008joint}
to obtain pronunciations for OOV terms. We find graphemes to be superior for Turkish and Zulu. This can be explained by the fact that when using phonemes, we only consider one pronunciation for each word, so the model learns a limited pronunciation model; whereas when we use graphemes, the model is allowed to implicitly learn pronunciation variability. However, for Pashto, phonemes work better. We hypothesize that this is because the graphemic representation uses the non-diacritized Pashto alphabet which omits vowels.

Finally, we utilize our system (referred to as \texttt{base} in Figure~\ref{fig:standalone}) to rescore the output of an LVCSR-based KWS system (multilingually pre-trained and finetuned on each target language, with words for IV and subwords for OOV search). We rescore using the procedure described in Section~\ref{subsec:rescoring} followed by KST normalization. The results in Table~\ref{tab:fusion} show that we are able to significantly improve on an already competitive system. 

%% file: figures/ablation.tex
\begin{axis}  
[  
    ybar,
    bar width=7pt,
    enlargelimits=0.2,
    legend style={at={(0.45,-0.15)},anchor=north,legend columns=5},
    ylabel={ATWV},
    symbolic x coords={Pashto, Turkish, Zulu},  
    xtick=data,
    ytick={0.05,0.1,0.15,0.2},
    yticklabel style={
        /pgf/number format/fixed,
        /pgf/number format/precision=2
    },
    width=0.46\textwidth,
    height=0.30\textwidth,
    nodes near coords align={vertical},
]
\addplot coordinates {(Pashto, 0.13) (Turkish, 0.1784) (Zulu, 0.1728)};
\addplot coordinates {(Pashto, 0.0439) (Turkish, 0.0791) (Zulu, 0.0612)};

\addplot coordinates {(Pashto, 0.0887) (Turkish, 0.1450) (Zulu, 0.1161)};
\addplot coordinates {(Pashto, 0.1067) (Turkish, 0.1606) (Zulu, 0.1722)};
\addplot coordinates {(Pashto, 0.1441) (Turkish, 0.1614) (Zulu, 0.1675)};
\legend{base,fbank,no-kst,no-margin,phoneme}
\end{axis}  

%% file: inputfiles/proclusion.tex
In this work, we have proposed a model that is directly optimized for keyword search. By predicting frame-wise likelihoods of keyword occurrences, the proposed model facilitates fine-grained keyword search and readily accommodates fusion with other systems.

We showed that our method works well not only for utterance classification used as a proxy for search but also for the more difficult ``needle-in-a-haystack" search problem - an encouraging result for the prospects of end-to-end approaches to keyword search. Moreover, we have shown that an application of our technique can already significantly improve LVCSR-based keyword search performance through rescoring.

%% file: main.bbl
\begin{thebibliography}{10}
\providecommand{\url}[1]{#1}
\csname url@samestyle\endcsname
\providecommand{\newblock}{\relax}
\providecommand{\bibinfo}[2]{#2}
\providecommand{\BIBentrySTDinterwordspacing}{\spaceskip=0pt\relax}
\providecommand{\BIBentryALTinterwordstretchfactor}{4}
\providecommand{\BIBentryALTinterwordspacing}{\spaceskip=\fontdimen2\font plus
\BIBentryALTinterwordstretchfactor\fontdimen3\font minus
  \fontdimen4\font\relax}
\providecommand{\BIBforeignlanguage}[2]{{%
\expandafter\ifx\csname l@#1\endcsname\relax
\typeout{** WARNING: IEEEtran.bst: No hyphenation pattern has been}%
\typeout{** loaded for the language `#1'. Using the pattern for}%
\typeout{** the default language instead.}%
\else
\language=\csname l@#1\endcsname
\fi
#2}}
\providecommand{\BIBdecl}{\relax}
\BIBdecl

\bibitem{can2011lattice}
D.~Can and M.~Sara{\c{c}}lar, ``Lattice indexing for spoken term detection,''
  \emph{IEEE Transactions on Audio, Speech, and Language Processing}, vol.~19,
  no.~8, pp. 2338--2347, 2011.

\bibitem{wang2008comparison}
D.~Wang, J.~Frankel, J.~Tejedor, and S.~King, ``A comparison of phone and
  grapheme-based spoken term detection,'' in \emph{ICASSP}, 2008, pp.
  4969--4972.

\bibitem{parlak2012performance}
S.~Parlak and M.~Sara{\c{c}}lar, ``Performance analysis and improvement of
  {Turkish} broadcast news retrieval,'' \emph{IEEE Transactions on Audio,
  Speech, and Language Processing}, vol.~20, no.~3, pp. 731--741, 2012.

\bibitem{garcia2006keyword}
A.~Garcia and H.~Gish, ``Keyword spotting of arbitrary words using minimal
  speech resources,'' in \emph{ICASSP}, vol.~1, 2006, pp. 949--952.

\bibitem{lee2005combining}
S.-w. Lee, K.~Tanaka, and Y.~Itoh, ``Combining multiple subword representations
  for open-vocabulary spoken document retrieval,'' in \emph{ICASSP}, vol.~1,
  2005, pp. 505--508.

\bibitem{mamou2007vocabulary}
J.~Mamou, B.~Ramabhadran, and O.~Siohan, ``Vocabulary independent spoken term
  detection,'' in \emph{Proceedings of the 30th annual international ACM SIGIR
  conference on Research and development in information retrieval}.\hskip 1em
  plus 0.5em minus 0.4em\relax ACM, 2007, pp. 615--622.

\bibitem{karakos2014subword}
D.~Karakos and R.~Schwartz, ``Subword and phonetic search for detecting
  out-of-vocabulary keywords,'' in \emph{Interspeech}, 2014, pp. 2469--2473.

\bibitem{szoke2008hybrid}
I.~Sz{\"o}ke, M.~Fap{\v{s}}o, and L.~Burget, ``Hybrid word-subword decoding for
  spoken term detection,'' in \emph{The 31st Annual International ACM SIGIR
  Conference on Research and Development in Information Retrieval}, 2008, pp.
  42--48.

\bibitem{saraclar2013confusion}
M.~Sara{\c{c}}lar, A.~Sethy, B.~Ramabhadran, L.~Mangu, J.~Cui, X.~Cui,
  B.~Kingsbury, and J.~Mamou, ``An empirical study of confusion modeling in
  keyword search for low resource languages,'' in \emph{{IEEE} Workshop on
  Automatic Speech Recognition and Understanding (ASRU)}, 2013, pp. 464--469.

\bibitem{chen2013using}
G.~Chen, O.~Yilmaz, J.~Trmal, D.~Povey, and S.~Khudanpur, ``Using proxies for
  oov keywords in the keyword search task,'' in \emph{2013 IEEE Workshop on
  Automatic Speech Recognition and Understanding}.\hskip 1em plus 0.5em minus
  0.4em\relax IEEE, 2013, pp. 416--421.

\bibitem{gundougdu2017joint}
B.~G{\"u}ndo{\u{g}}du, B.~Yusuf, and M.~Sara{\c{c}}lar, ``Joint learning of
  distance metric and query model for posteriorgram-based keyword search,''
  \emph{IEEE Journal of Selected Topics in Signal Processing}, vol.~11, no.~8,
  pp. 1318--1328, 2017.

\bibitem{yusuf2019low}
B.~Yusuf, B.~Gundogdu, and M.~Saraclar, ``Low resource keyword search with
  synthesized crosslingual exemplars,'' \emph{IEEE/ACM Transactions on Audio,
  Speech, and Language Processing}, vol.~27, no.~7, pp. 1126--1135, 2019.

\bibitem{chen2014small}
G.~Chen, C.~Parada, and G.~Heigold, ``Small-footprint keyword spotting using
  deep neural networks,'' in \emph{ICASSP}.\hskip 1em plus 0.5em minus
  0.4em\relax IEEE, 2014, pp. 4087--4091.

\bibitem{Segal2019}
\BIBentryALTinterwordspacing
Y.~Segal, T.~S. Fuchs, and J.~Keshet, ``{SpeechYOLO: Detection and Localization
  of Speech Objects},'' in \emph{Proc. Interspeech 2019}, 2019, pp. 4210--4214.
  [Online]. Available: \url{http://dx.doi.org/10.21437/Interspeech.2019-1749}
\BIBentrySTDinterwordspacing

\bibitem{palaz2016jointly}
D.~Palaz, G.~Synnaeve, and R.~Collobert, ``Jointly learning to locate and
  classify words using convolutional networks,'' \emph{Interspeech}, pp.
  2741--2745, 2016.

\bibitem{alvarez2019end}
R.~Alvarez and H.-J. Park, ``End-to-end streaming keyword spotting,'' in
  \emph{ICASSP}.\hskip 1em plus 0.5em minus 0.4em\relax IEEE, 2019, pp.
  6336--6340.

\bibitem{audhkhasi2017end}
K.~Audhkhasi, A.~Rosenberg, A.~Sethy, B.~Ramabhadran, and B.~Kingsbury,
  ``End-to-end asr-free keyword search from speech,'' \emph{IEEE Journal of
  Selected Topics in Signal Processing}, vol.~11, no.~8, pp. 1351--1359, 2017.

\bibitem{chen2019audio}
Y.-C. Chen, S.-F. Huang, H.-y. Lee, Y.-H. Wang, and C.-H. Shen, ``Audio
  word2vec: Sequence-to-sequence autoencoding for unsupervised learning of
  audio segmentation and representation,'' \emph{IEEE/ACM Transactions on
  Audio, Speech, and Language Processing}, vol.~27, no.~9, pp. 1481--1493,
  2019.

\bibitem{kamper2020improved}
H.~Kamper, Y.~Matusevych, and S.~Goldwater, ``Improved acoustic word embeddings
  for zero-resource languages using multilingual transfer,'' \emph{IEEE/ACM
  Transactions on Audio, Speech, and Language Processing}, vol.~29, pp.
  1107--1118, 2021.

\bibitem{zhao2020end}
Z.~Zhao and W.-Q. Zhang, ``End-to-end keyword search based on attention and
  energy scorer for low resource languages,'' in \emph{Interspeech}, 2020, pp.
  2587--2591.

\bibitem{harper2014iarpa}
M.~Harper, ``{IARPA} {B}abel program,'' \url{https://www. iarpa. gov/index.
  php/research-programs/babel}, 2014, accessed at March 2021.

\bibitem{Fiscus2006}
J.~Fiscus, J.~Ajot, and G.~Doddington, ``The spoken term detection (std) 2006
  evaluation plan,'' \emph{NIST USA}, 2006.

\bibitem{Fiscus2007}
J.~G. Fiscus, J.~Ajot, J.~S. Garofolo, and G.~Doddingtion, ``{Results of the
  2006 spoken term detection evaluation},'' in \emph{Proceedings of the ACM
  SIGIR Workshop on Searching Spontaneous Conversational Speech}, 2007, pp.
  51--57.

\bibitem{kws14Evalplan}
``Open{KWS}14 keyword search evaluation plan,''
  \url{http://www.nist.gov/itl/iad/mig/upload/KWS14-evalplan-v11.pdf}, accessed
  at March 2021.

\bibitem{byers20192017}
F.~Byers and O.~Sadjadi, \emph{2017 Pilot Open Speech Analytic Technologies
  Evaluation (2017 NIST Pilot OpenSAT): Post Evaluation Summary}.\hskip 1em
  plus 0.5em minus 0.4em\relax US Department of Commerce, National Institute of
  Standards and Technology, 2019.

\bibitem{miller2007rapid}
D.~R. Miller, M.~Kleber, C.-L. Kao, O.~Kimball, T.~Colthurst, S.~A. Lowe, R.~M.
  Schwartz, and H.~Gish, ``Rapid and accurate spoken term detection,'' in
  \emph{Interspeech}, 2007, pp. 314--317.

\bibitem{bisani2008joint}
M.~Bisani and H.~Ney, ``Joint-sequence models for grapheme-to-phoneme
  conversion,'' \emph{Speech communication}, vol.~50, no.~5, pp. 434--451,
  2008.

\end{thebibliography}
